\documentclass{PoS}

\title{Study of two- and three-meson decay modes of tau-lepton
with Monte Carlo generator TAUOLA}

\ShortTitle{Tau-lepton decays}

\author{\speaker{Olga Shekhovtsova}\thanks{IFJPAN-IV-2015-15}
\\
        Institute of Nuclear Physics PAN ul. Radzikowskiego 152 31-342 Krakow, Poland\\
 Kharkov Institute of Physics and Technology   61108, Akademicheskaya,1, Kharkov, Ukraine\\
        E-mail: \email{oshekhov@ifj.edu.pl}}

\abstract{
The study of the $\tau$-lepton decays into hadrons has contributed to a better understanding of non-perturbative QCD and light-quark meson spectroscopy,  as well as to the search of new physics beyond the Standard Model. 
The two- and three-meson decay modes, considering only those permitted by the Standard Model,  are the predominant decays and together with the one-pion mode compose more than $85\%$ of the hadronic $\tau$-lepton decay width. In this note   we review the theoretical results for these modes implemented in the Monte Carlo event generator TAUOLA  and present at the same time a comparison with the Belle Collaboration data for the two-pion decay mode and the BaBar preliminary data for the three-pion decay mode as  well for the decay mode into two-kaon and one-pion.
}

\FullConference{The 8th International Workshop on Chiral Dynamics, CD2015 ***\\
		29 June 2015 - 03 July 2015\\
		Pisa,Italy}

\begin{document}

\section{Introduction}\label{sect:intro}

Tau lepton is a fundamental particle of the Standard Model (SM) and knowledge of its properties with high accuracy is absolutely mandatory for precise SM tests~\cite{Pich:2013lsa}.
Its hadronic decay modes provide a unique laboratory to study and develop low energy QCD. They also allow to measure SM fundamental parameters like the QCD strong coupling, elements of the Cabibbo-Kobayashi-Maskawa matrix, the strange quark mass.  Also hadronic modes of $\tau$-lepton decays play a critical role as a probe to search for signals of new physics beyond SM~\cite{Celis:2013xja}.  

Since the 90's  TAUOLA~\cite{Jadach:1993hs} is the main Monte Carlo event generator that is applied to simulate $\tau$-lepton decay events in the analysis of experimental data both at B-factories and LHC.  It has been used by the collaborations ALEPH \cite{Buskulic:1995ty}, CLEO  \cite{Asner:1999kj}, at both B-factories (BABAR \cite{Nugent:2013ij} and BELLE~\cite{Fujikawa:2008ma}) as well at LHC \cite{LHC1,LHC2} experiments. The library TAUOLA can be easily attached to any MC describing the production mechanism like KORALB, KORALZ and KKMC \cite{koral1,koral2}. The code provides the full topology of the final particles including their spin. Currently the code is capable to simulate more than 20 hadronic decay modes. 

In this note we discuss the status of the SM two- and three-meson channels of the $\tau$-lepton decay installed in TAUOLA. Mainly, we concentrate on the $\pi^-\pi^0$, $\pi^+\pi^-\pi^-$ and $K^+K^-\pi^-$ decay modes.
Our choice is related with the fact that being the predominant two- and three-meson decay modes, these channels give information about resonances involved in production and about the hadronization mechanism~\cite{Bevan:2014iga}.  In addition, using the conservation of vector current and correcting for isospin-violating effects, the precise data for the two-pion mode can be used to estimate the leading-order hadronic contribution  to the anomalous magnetic momentum of the muon~\cite{Alemany:1997tn}. Also the three-pion channels
together with the two-pion decay mode are used for studies of the Higgs-lepton coupling at LHC and for the tau-lepton measurements.

\section{Two-meson decay modes of $\tau$-lepton }\label{sect:two_mes}
The MC TAUOLA contains altogether the two-meson decay modes into: two pions ($Br \simeq 25.52\%$), two kaons ($Br \simeq 0.16\%$), one pion and a kaon ($Br \simeq 1.27 \%$). Therefore the modes with the $\eta$($\eta'$) meson have not yet been included in the code. Motivations for the $\eta$ decay mode measurements as well the related hadronic current calculated in the framework of the  RChL approach can be found in \cite{Escribano:2014joa}.  

\subsection{Hadronic current of two-meson decay modes}
For $\tau$ decay channels with two mesons, $h_1(p_1)$ and $h_2(p_2)$ with masses $m_1$ and $m_2$, respectively, the hadronic current reads
\begin{equation}
J^\mu  = N \bigl[ (p_1 - p_2 -\frac{\Delta_{12}}{s}(p_1+p_2))^\mu F^{V}(s) + \frac{\Delta_{12}}{s}(p_1 + p_2)^\mu F^{S}(s) \bigr],
\end{equation}
where $s = (p_1 +p_2)^2$ and $\Delta_{12} = m_1^2 - m_2^2$. 
 The normalization factor $N$ is equal 1 for the $\pi^-\pi^0$ channel, while the other three normalization factors are related by SU(3) symmetry using the Clebsh-Gordan coefficients:
\begin{equation}
N^{K^- K^0} = \frac{1}{\sqrt{2}} \, , \; \; N^{\pi^-K^0} = \frac{1}{\sqrt{2}} \, , \; \; N^{\pi^0K^-} = \frac{1}{2} \, .
\end{equation}
The formulae for the vector ($F^{V}(s)$) and scalar ($F^{S}(s)$) form factors depends on the channel.  In the general case both vector and scalar form factors are present. In the isospin symmetry limit, $m_{\pi^\pm} = m_{\pi^0}$, $m_{K^\pm} = m_{K^0}$, the scalar form factor vanishes for both two-pion and two-kaon modes  and the corresponding channel is described by the vector form factor alone.

\subsection{Two-pion form factor and comparison with the Belle data}

As mentioned above the scalar factor contribution for the two-pion channel is negligibly small, so the width is defined by the pion vector form factor alone.
Currently MC TAUOLA includes four parametrizations for the vector form factor of two pions $F^{V}_{\pi}(s)$: 
\begin{itemize} 
\item Kuhn-Santamaria (KS) parametrization~\cite{Kuhn:1990ad}:
\begin{eqnarray*}
F^{V}_{\pi}(s) = \frac{1}{1+\beta+\gamma}(BW_\rho(s) +\beta BW_{\rho'}(s)+\gamma BW_{\rho''}(s)) \, , \,  \,  BW(s) = \frac{M^2}{M^2-s-i\sqrt{s}\Gamma_{\pi\pi}(s)} \, ,
\end{eqnarray*} 
where $M$ is a resonance mass and $\Gamma_{\pi\pi}(s)$ is the resonance energy-dependent width  that takes into account two-pion loops;
\item Gounaris-Sakurai (GS) parametrization used by BELLE~\cite{Fujikawa:2008ma}, ALEPH and CLEO collaborations: 
\begin{eqnarray*}
F^{V}_{\pi}(s) &=& \frac{1}{1+\beta+\gamma}(BW_\rho^{GS}(s) +\beta BW_{\rho'}^{GS}(s)+\gamma BW_{\rho''}^{GS}(s)) \, , \\ BW^{GS}(s) &=& \frac{M^2+d M\Gamma_{\pi\pi}(s)}{M^2-s+f(s)-i\sqrt{s}\Gamma_{\pi\pi}(s)} \, , 
\end{eqnarray*} 
where $f(s)$ includes the real part of the two-pion loop function;
\item parametrization based on the Resonance Chiral Lagrangian (RChL)~\cite{SanzCillero:2002bs}:
\begin{equation}
F^{V}_{\pi}(s) = \frac{1+\sum\limits_{i = \rho, \rho', \rho''}\frac{F_{V_i}G_{V_i}}{F^2}\frac{s}{M^2_{i}-s}}
                     {1+\left(1+\sum\limits_{i = \rho, \rho',\rho''} \frac{ 2G^2_{V_i}}{F^2}\frac{s}{M^2_{i}-s}\right)\frac{2s}{F^2}\left[B_{22}^{\pi}(s)+\frac{1}{2}B_{22}^{K}(s)\right]} \, , \nonumber
\end{equation}
where $B_{22}$ is the two-meson loop function~\footnote{Comparing the imaginary part of the loop function $B_{22}^{(\pi)}$, Eq.(A.3) in Ref.~\cite{SanzCillero:2002bs}, and Eq. (13) in~\cite{Fujikawa:2008ma} one obtains $\sqrt{s}\Gamma_{\pi\pi}(s)= s\sqrt{M_V}\Gamma_V \displaystyle\frac{{\rm Im} B_{22}^\pi(s)}{{\rm Im} B_{22}^\pi(M_V)}$ for $s>(m_{\pi^-}+m_{\pi^0})^2$, where $M_i$ and $\Gamma_i$ are the resonance mass and width, respectively.}. For the physical meaning of the model parameters $F_{V_i}$ and $G_{V_i}$ see~\cite{SanzCillero:2002bs,Ecker:1988te};
\item combined parametrization (combRChL) that applies dispersion approximation at low energy and modified RChL result at high energy~\cite{Dumm:2013zh}:
\begin{eqnarray*}
s < s_0 \,:  \; \; \; \; F^{V}_{\pi}(s) &=& \exp\left[\alpha_1 s + \frac{\alpha_2}{2}s^2 + \frac{s^3}{\pi}\int\limits_{4m_\pi^2}^{\infty} ds'\frac{\delta_1^1(s')}{(s')^3(s'-s-i\epsilon)} \right] \, ,   \\
s > s_0 \, : \; \; \; \; F^{V}_{\pi}(s) &=& \frac{M_\rho^2 +(\beta + \gamma)s}{M_\rho^2 -s+\displaystyle\frac{2s}{F^2}M_\rho^2\left[B_{22}^{\pi}(s)+\frac{1}{2}B_{22}^{K}(s)\right]}
   \\
&-&\frac{\beta s}{M^2_{\rho'}-s + \displaystyle\frac{192\pi s\Gamma_{\rho'}}{M_{\rho'}\sigma_\pi^3}B_{22}^\pi(s)} -
\frac{\gamma s}{M^2_{\rho''}-s + \displaystyle\frac{192\pi s\Gamma_{\rho''}}{M_{\rho''}\sigma_\pi^3}B_{22}^\pi(s)}  ,
\end{eqnarray*} 
where $s_0$ is the high energy limit of the applicability of the dispersion representation. It is supposed to satisfy $1.0$GeV$^2 < s_0 < 1.5$GeV$^2$~\cite{Dumm:2013zh} and we leave it as a fitting parameter. 
\end{itemize}  

In all the above parametrizations, except for the RChL one, the pion form factor is given by interfering amplitudes from the known isovector meson resonances $\rho(770)$, $\rho'(1450)$ and $\rho''(1700)$ with relative strengths $1$, $\beta$ and $\gamma$.  Although one could expect from the quark model that $\beta$ and $\gamma$ be real, we allow these parameters to be complex (following the BELLE, CLEO and ALEPH analysis) with their phases are left free in the fits. 
In the case of the RChL parametrization we restrict ourselves to  the $\rho(700)$ and $\rho'(1450)$ contributions, the relative $\rho'$ strength  (which is a combination of the model parameters $F_{V_i}$, $G_{V_i}$ and $F$) being a real parameter. 
 
For the energy-dependent width  of the $\rho(770)$-meson, two-pion and two-kaon loop contributions are included for both  RChL and combRChl parametrizations, whereas in the case of KS and GS the $\rho$ width is approximated only by the two-pion loops. 
The $\rho'(1450)$ and $\rho''(1700)$ widths include only two-pion loops for all parametrizations except for the RChL one. In this later case both two-pion and two-kaon loops are included.

Results of the fit to the BELLE data \cite{Fujikawa:2008ma} are presented in Figs.~1 and 2. The best fit is obtained with the GS pion form factor ($\chi^2 = 95.65 $) and the worst with the RChL one ($\chi^2 = 156.93 $), that is not able to reproduce the high energy tail. As mentioned above, the two main differences of the RChL parametrization compared to the others are 1) the $\rho''$-meson absence, 2) a real value of the $\rho'(1450)$-meson strength. To check the influence of the $\rho''(1700)$ on the RChL result, this resonance has been included in the same way as for $\rho'(1450)$; however, this inclusion has not improved the result (see, Fig. 2, right panel).

In an effective field theory, like RChL, complex values come only from loops.
Therefore, we conclude that missing loop contributions could be responsible for the disagreement and that the complex value of the $\beta$ and $\gamma$ parameters might mimic  missing multiparticle loop contributions. The same conclusion was reached in Ref.~\cite{SanzCillero:2002bs} where it was stressed that the two-pseudoscalar loops cannot incorporate all the inelasticity needed
to describe the data and other multiparticle intermediate states can play a role. This point will be checked by adding first a four-pion loop contribution
 to the $\rho'$-resonance propagator.
\begin{figure}[h]
\includegraphics[scale = .32]{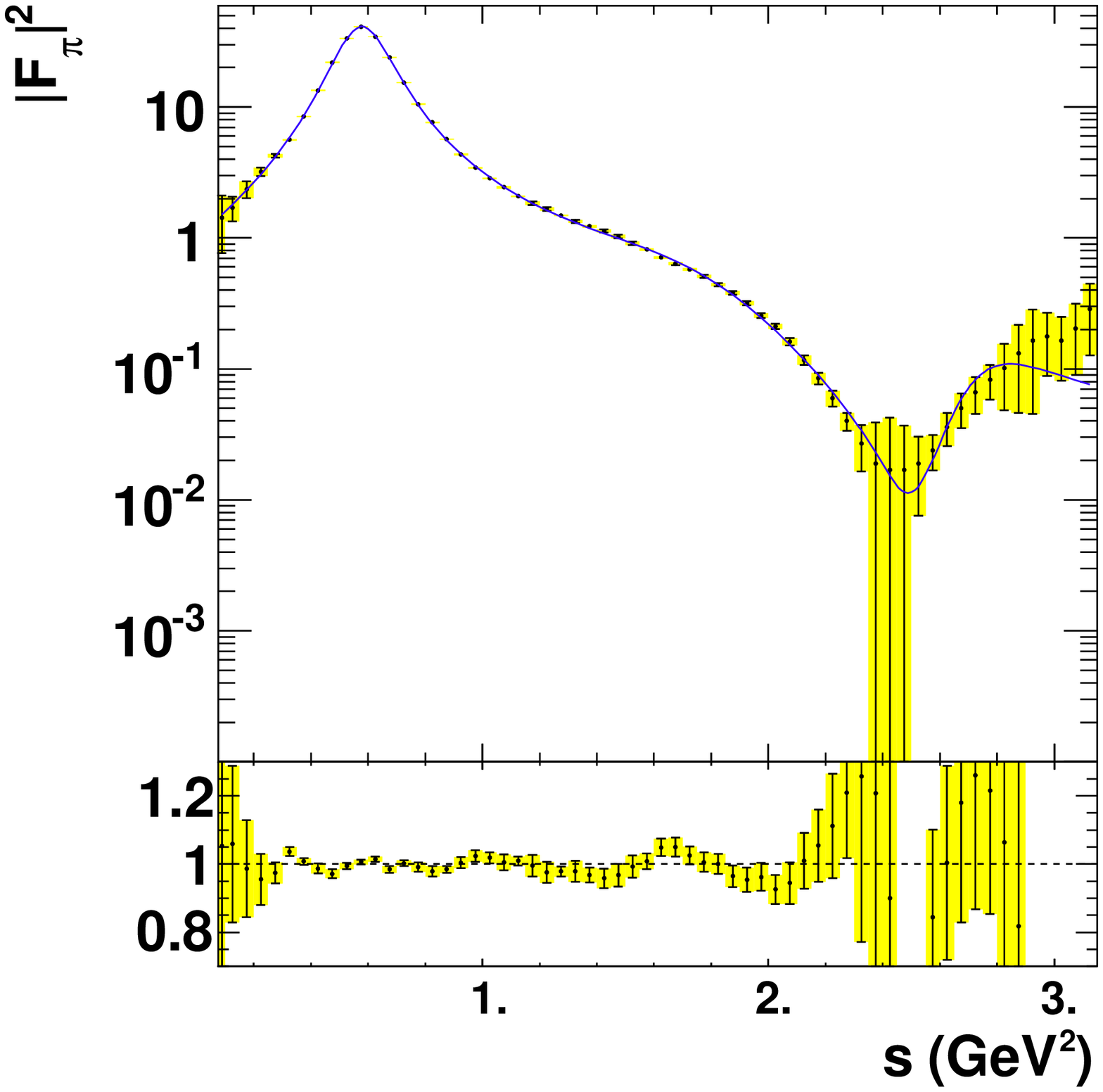}
\hspace{0.8cm}
\includegraphics[scale = .32]{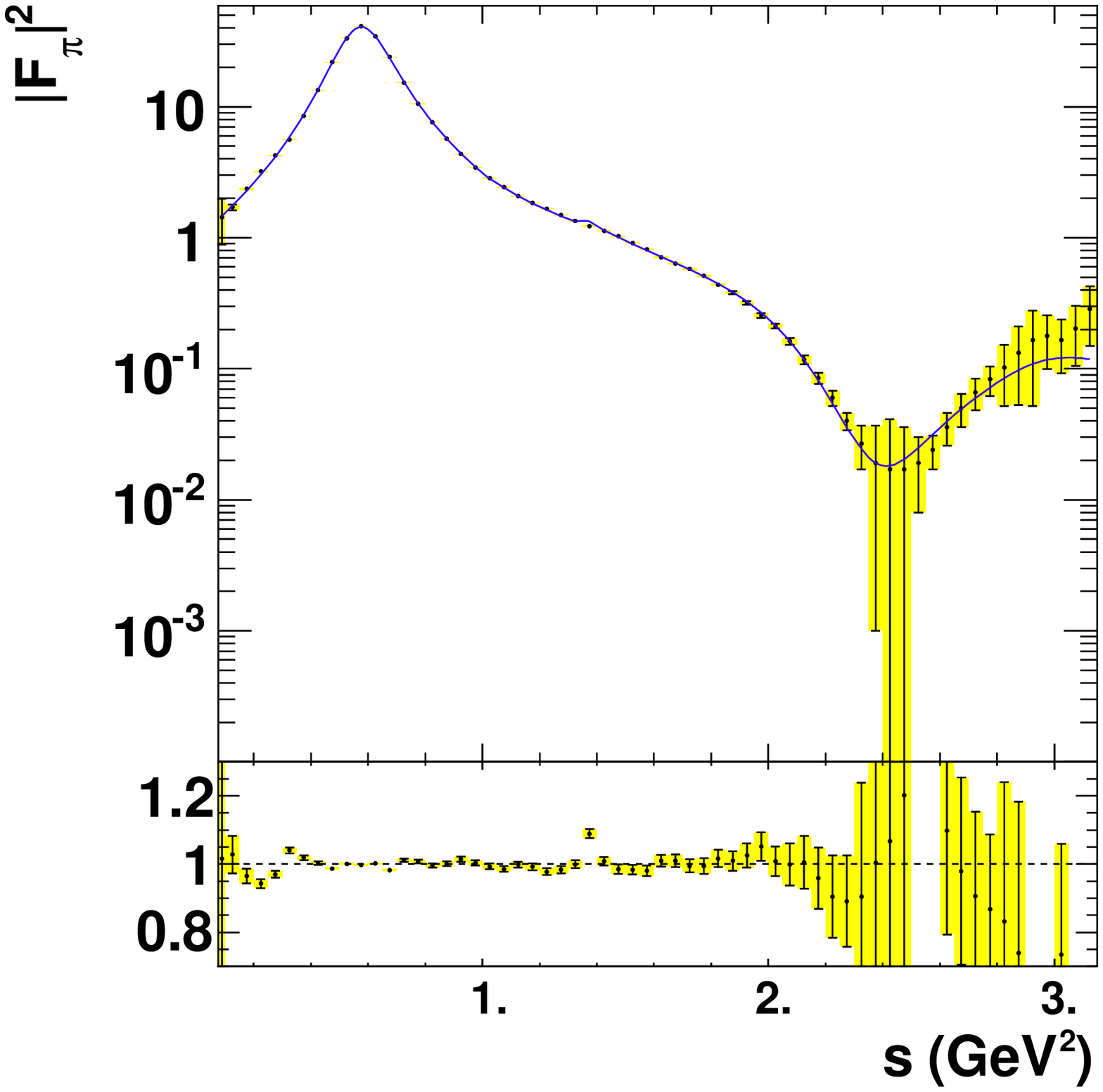}
\caption{The pion form factor fit to Belle data \cite{Fujikawa:2008ma}: the GS parametrization (left panel), the RChL parametrization (right panel). At the bottom of the figure, the ratio of the theoretical prediction
to the data is given.}
\end{figure}

\begin{figure}[h]
\includegraphics[scale = .32]{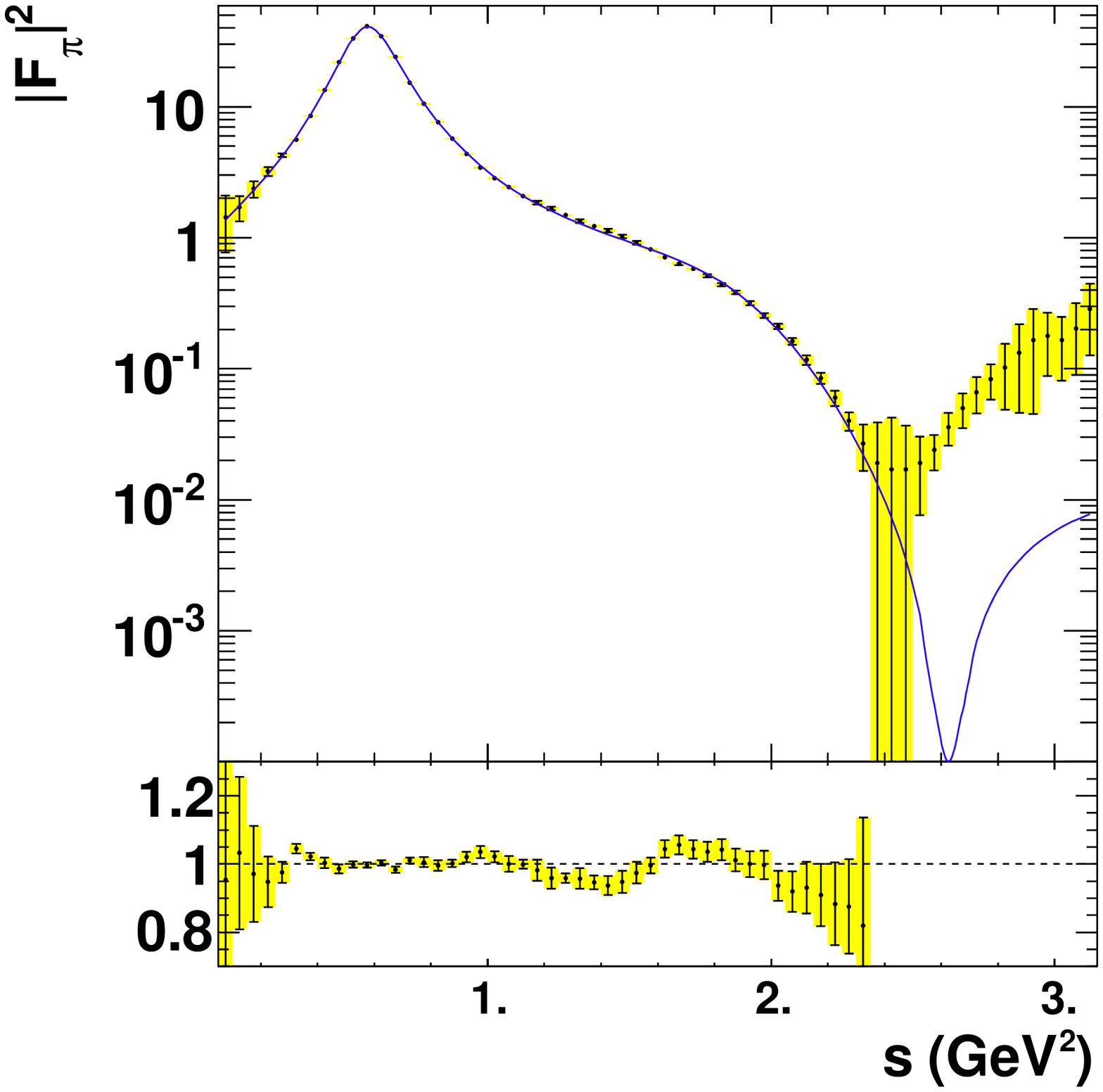}
\hspace{0.8cm}
\includegraphics[scale = .32]{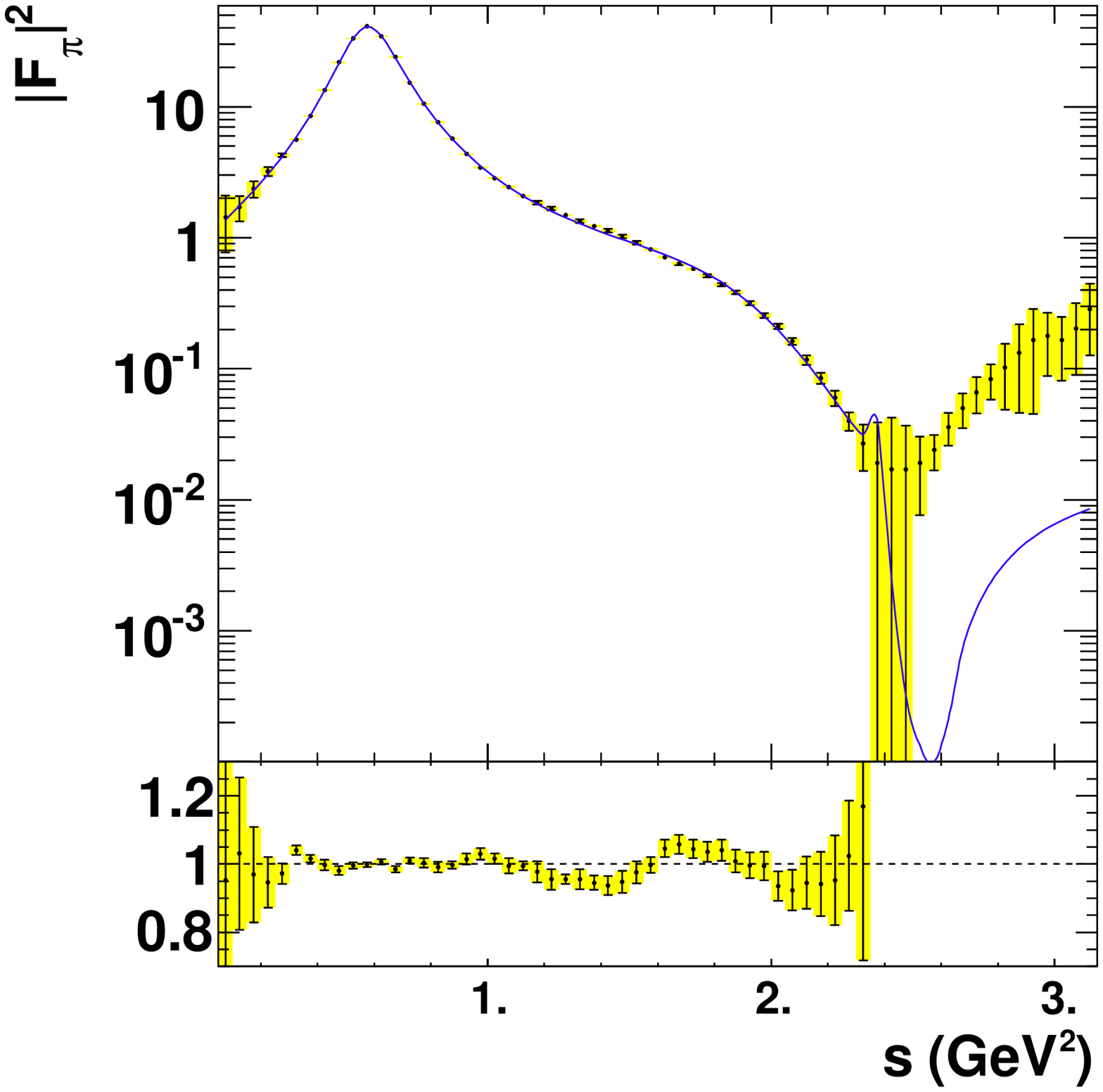}
\caption{The pion form factor within the comRChL parametrization is fitted to BELLE data \cite{Fujikawa:2008ma}: the fit without $\rho''(1700)$ (left panel) and with it (right panel). At the bottom of the figure, the ratio of the theoretical prediction
to the data is given.}
\end{figure}


In the case of the combRChL parametrization the fitting curvature does not present a smooth behaviour near $s = s_0$. Therefore  more sophisticated fitting techniques will have to be implemented.

\subsection{Two-kaon and kaon-pion decay modes}
The expressions for the  two-kaon vector form factor coincide with the two-pion form factor. Currently, we have implemented only one parametrization in the TAUOLA code, namely the  modified RChL result~Eq.(26) of Ref.~\cite{Shekhovtsova:2012ra}. Assuming the $SU(3)$ symmetry we have kept the same value for the parameters $\gamma$ and $\delta$ and estimated a partial width $(2.65\pm 0.01\%)\dot 10^{-15}$ GeV, that is only about $~60\%$ of the PDG value~\cite{Agashe:2014kda}. As corrections of order $~30\%$ are expected it would be interesting to make a direct fit of the two-kaon form factor to the corresponding experimental data. However, till now only the branching ratio is available~\cite{Ryu:2014vpc}. 

The $K \pi$ decay mode measurement allows to measure the $K^*$-resonance parameters as well the Cabibbo-Kabayshi-Maskawa matrix elemenet $|V_{us}|$. For this mode both scalar and vector form factors play a role.
Currently, the only the following parametrizations of the vector kaon-pion form factor have been implemented in TAUOLA: the RChL approach~\cite{Jamin:2008qg} and the parametrization based on the dispersion approximation~\cite{Boito:2008fq}. The scalar form factor is computed using the private code of M. Jamin~\cite{Jamin:2001zq}. The model parameters are fixed to their values from~\cite{Jamin:2008qg,Boito:2008fq} and a fit to the Belle data will be a task of your future work.

\section{Three-meson decay modes of $\tau$-lepton}
The following three-meson decay channels are implemented in TAUOLA~\cite{Jadach:1993hs,Shekhovtsova:2012ra}: three-pion ($\pi^0\pi^0\pi^-$ and $\pi^-\pi^-\pi^+$) modes, two-kaon and one-pion ($K^-\pi^-K^+$, $K^0\pi^0\bar{K}^0$, $K^-\pi^0K^0$), two-pion and one-kaon ($K^-\pi^0\pi^0$, $K^-\pi^-\pi^+$, $K^0\pi^0\pi^-$), two-pion and the $\eta$-meson ($\eta\pi^0\pi^-$). Two theoretical parametrizations for the hadronic form factors, the RChL approach and the standard VMD approximation, have been implemented in the code, whereas the other channels are based only on the VMD approximation~\cite{Jadach:1993hs}.   

\subsection{Three-meson hadronic currents and form-factors}
For the final state of three pseudoscalars, with momenta $p_1$, $p_2$,$p_3$ and masses $m_1$, $m_2$, $m_3$, respectively, the most general hadronic current compatible with the Lorentz invariance can be written as 
\begin{eqnarray}
J^\mu &=N &\bigl\{T^\mu_\nu \bigl[ c_1 (p_2-p_3)^\nu F_1(q^2,s_1,s_2)  + c_2 (p_3-p_1)^\nu
 F_2(q^2,s_1,s_2)  + c_3  (p_1-p_2)^\nu F_3(q^2,s_1,s_2) \bigr]\nonumber\\
& & + c_4  q^\mu F_4(q^2,s_1,s_2)  -{ i \over 4 \pi^2 F^2}      c_5
\epsilon^\mu_{.\ \nu\rho\sigma} p_1^\nu p_2^\rho p_3^\sigma F_5(q^2,s_1,s_2)      \bigr\},
\label{eq:curr_3mes}
\end{eqnarray}
where as usual  $T_{\mu\nu} = g_{\mu\nu} - q_\mu q_\nu/q^2$ denotes the transverse
projector, $q^\mu=(p_1+p_2+p_3)^\mu$ is the total momentum of the hadronic system  and the two-meson invariant mass squared is given by $s_i = (p_j + p_k)^2$. Here and afterward in the paper $F$ stands for the pion decay constant in the chiral limit. The normalization coefficient is $N = \cos{\theta_{Cabibbo}}$ for modes with an even kaon numbers, otherwise $N = \sin{\theta_{Cabibbo}}$.

The scalar functions $F_i(q^2,s_1,s_2)$ are the hadronic form factors. In general they depend on three independent invariant
masses that can be constructed from the three meson four-vectors:  we chose $q^2$, $s_1$, $s_2$. Of the hadronic form factors $F_i$, $i = 1, 2, 3$  which correspond 
to the axial-vector part of the hadronic tensor, only two are independent, however for convenience we keep all of them in Eq.~(\ref{eq:curr_3mes}) and in the code. The pseudoscalar form factor $F_4$ is proportional to $m^2_\pi/q^2$~\cite{Shekhovtsova:2012ra}, thus it is suppressed with
respect to $F_i$, $i = 1, 2, 3$. The vector form factor vanishes for the three-pion modes due to the
G-parity conservation: $F_5^{3\pi} = 0$. 

\subsection{Comparison with the BaBar preliminary data for the $\pi^+\pi^-\pi^-$ decay mode}

Among the three-meson $\tau$-lepton decay channels  the three-pion modes have the largest value, $Br \simeq 9.3\%$ for $\pi^0\pi^0\pi^-$ and $Br \simeq 9.0\%$ for $\pi^+\pi^-\pi^-$.  We would like to remind that precise modeling of the three-pion modes are important not only for the study of the hadronization in itself but also for the tau-lepton mass measurement and, together with two-pion decay mode, it is used for studies of the Higgs-lepton coupling by Alice and CMS Collaborations at LHC, Cern.

In TAUOLA the following three-pion form factors are available 
\begin{itemize} 
\item CPC version~\cite{Jadach:1993hs}, that includes only the dominant $a_1 \to \rho\pi$ mechanism production. The form factor is a product of the Breit-Wigner amplitudes for the $a_1$ and $\rho$ mesons;  
\item CLEO parametrization is based on the Dalitz plot analysis carried out by the CLEO collaboration and includes the following intermediate states: $a_1 \to (\rho;\rho')\pi$, $a_1 \to \sigma\pi$, $a_1 \to f_2(1270)\pi$, $a_1 \to f_0(1370)\pi$.   
In fact, there are two variants of this parametrization. The former is based on the CLEO  $\pi^0\pi^0\pi^-$  analysis  \cite{Asner:1999kj} and  applies the same current for the $\pi^-\pi^-\pi^+$ mode.
 The latter uses the $\pi^0\pi^0\pi^-$ current from \cite{Asner:1999kj} and the $\pi^-\pi^-\pi^+$ current from  the unpublished CLEO analysis~\cite{Shibata:2002uv}. \footnote{It is interesting to point out that the difference between these variants of the CLEO parametrization is related with the scalar and tensor resonance contributions. More recent discussion on this topic can be found in~\cite{Was:2015laa}.} All resonances are modeled by Breit-Wigner functions and the hadronic current is a weighted sum of their product~\footnote{This approach was contested in Ref.~\cite{GomezDumm:2003ku} where it was demonstrated that the corresponding hadronic form factors reproduced the leading-order chiral result and failed to reproduce
the next-to-leading-order one.}. The model parameters are the resonance masses and widths as well as their weights;
\item modified RChL parametrization \cite{Nugent:2013hxa}. 
It is based on the RChl results for the three-pion currents~\cite{Dumm:2009va} and an additional scalar resonance contribution. The RChL current is a sum of the chiral contribution corresponding to the direct vertex $W^- \to \pi\pi\pi$, single-resonance contributions, e.g. $W^- \to \rho \pi$, double-resonance contributions, as $W^- \to a_1^- \to \rho \pi$. Only vector and axial-vector are included in the RChL hadronic currents. The scalar resonance contribution was included phenomenologically by requiring the RChL structure for the currents and modelling the $\sigma$-resonance by a  Breit-Wigner function.
It is worth mentioning that the main numerical problem was related with the $a_1$-resonance width. The $a_1$-width entering the $a_1$-resonance propagator, is written down as the imaginary
part of the two-loop axial-vector-axial-vector correlator~\cite{Dumm:2009va} and is a double integral  of the same hadronic form factors that appear in the hadronic currents (for details, see \cite{Shekhovtsova:2012ra}, Section 3). We apply the 16-point Gaussian quadrature method to make the corresponding double integrations. 
More details about the modified RChL parametrization can be found in~\cite{Shekhovtsova:2012ra}.
\end{itemize} 

The CLEO parametrization for the $\pi^-\pi^-\pi^+$ mode has not yet been fitted to the BABAR preliminary data \cite{Nugent:2013ij}, so, for comparison with the BaBar preliminary data and the prediction based on the modified RChL parametrization, we use the numerical values of the parameters fitted to the old CLEO data~\cite{Asner:1999kj}. The fit to the BABAR data will be a task for future work.
\begin{figure}[h]
\includegraphics[scale = .24]{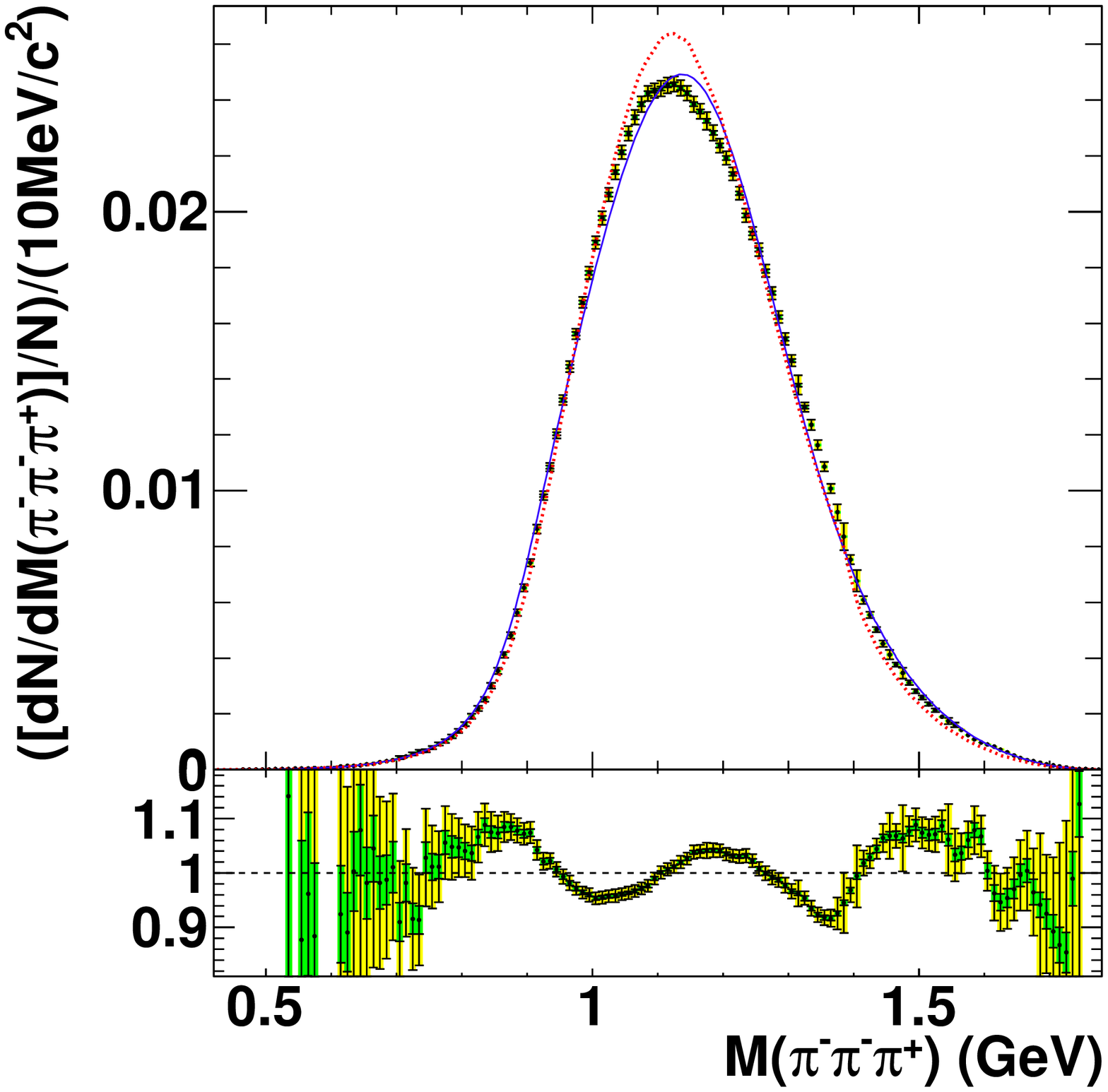}
\includegraphics[scale = .24]{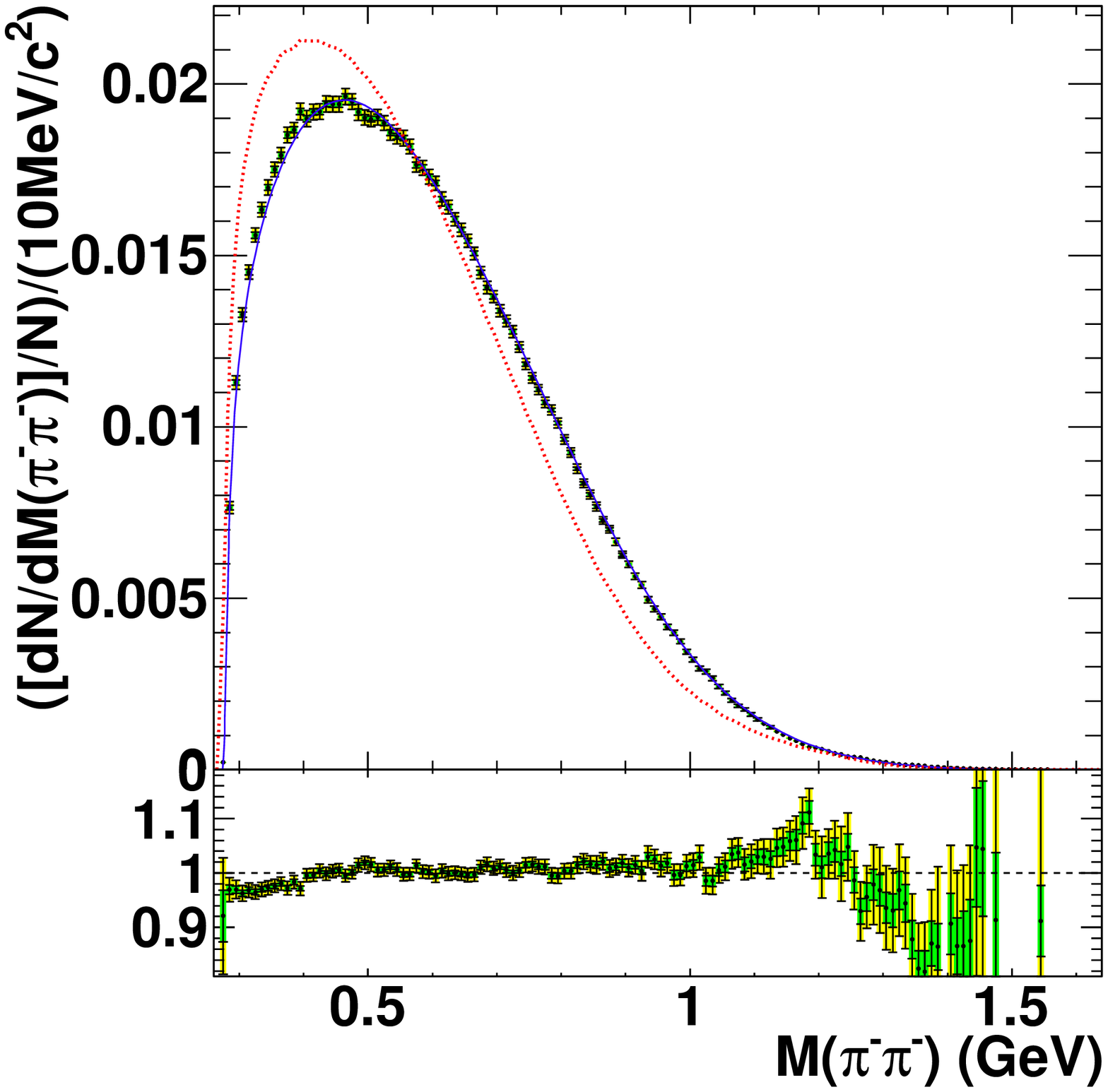}
\includegraphics[scale = .24]{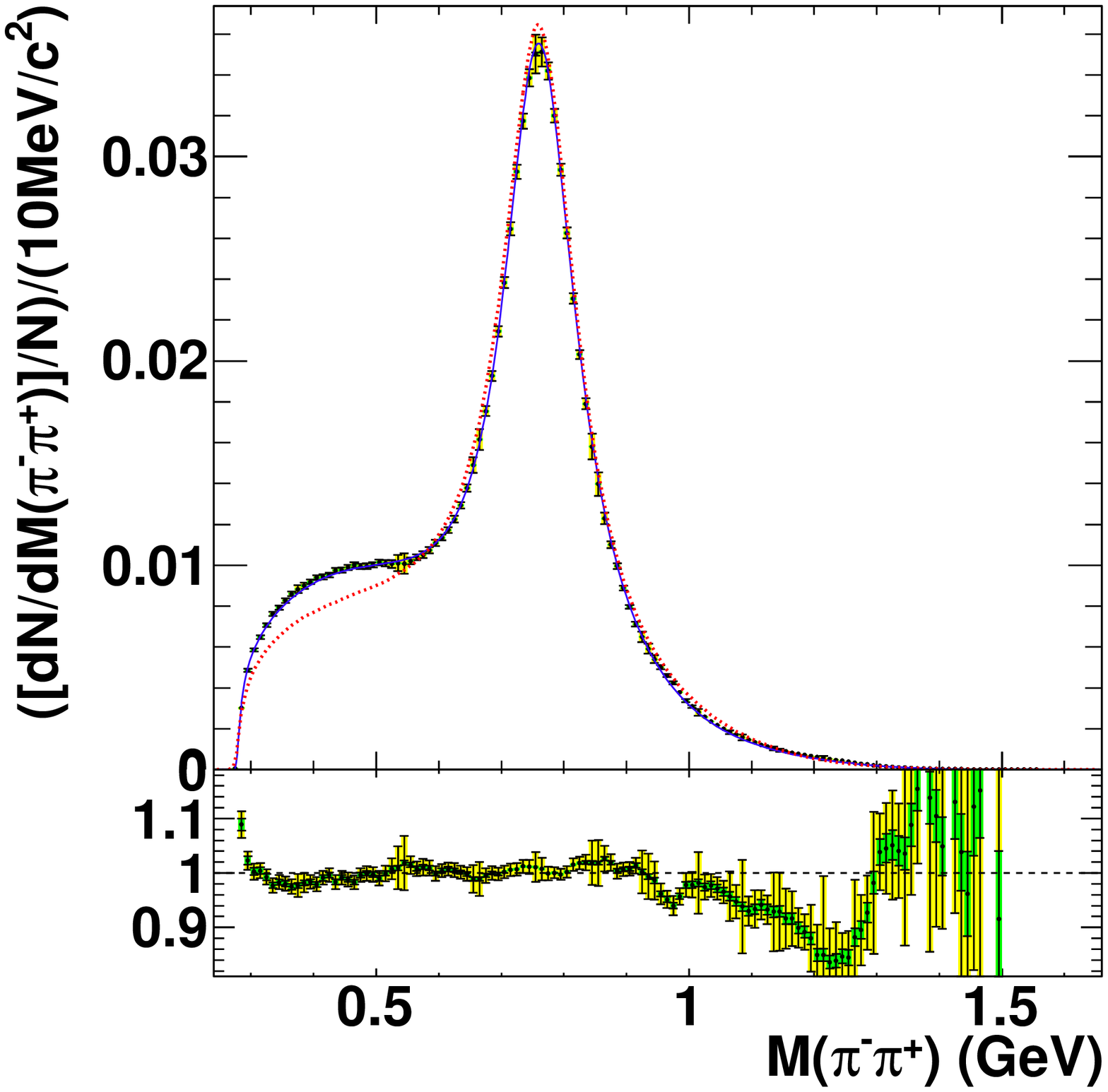}
\caption{The $\tau^-\to \pi^-\pi^-\pi^+\nu_\tau$ decay invariant mass distribution of the three-pion system (left panel) and the two-pion system (central and right panels).
The BABAR data~\cite{Nugent:2013ij} are represented by the data points, overlaid by the results from the modified RChL
current as described in the text (blue line) and the old fit curve from CLEO~\cite{Asner:1999kj} 
(red-dashed line) overlaid.
}
\end{figure}

The one-dimensional distributions of the two- and three-pion invariant mass spectra calculated on the base of the modified RChL parametrization have been fitted to the BABAR preliminary data~\cite{Nugent:2013ij}. 
Finally, after the introduction of the parallelized calculation, the precise calculations of the $a_1$ width value was incorporated into the project and its value is recalculated at every step of the fit iteration.
It must be pointed out that without the scalar resonance contribution the RChL parametrization provides a slightly better result than the CLEO parametrization whereas the scalar resonance inclusion strongly improves the low two-pion mass invariant spectrum.
Discrepancy between  theoretical spectra and experimental data can be explained by missing resonances in the model, such as the axial-vector resonance $a_1'(1600)$, the scalar resonance $f_0(980)$ and the tensor resonance $f_2(1270)$. Inclusion of these resonances in the RChL framework will be a future task.

Comparison of the $\pi^-\pi^-\pi^+$ current in the framework of the modified RChL with the ChPT result has demonstrated that the scalar resonance contribution has to be corrected to reproduce the low energy ChPT limit. The corresponding calculation is in progress.

\subsection{Comparison with the BaBar preliminary data for the $K^+K^-\pi^-$ decay mode}

Contrary to the three-pion channels the decay  $\tau^- \to K^+K^-\pi^- \nu_\tau$ depends  both on the vector and axial vector currents. 
Two parametrizations for the hadronic form factors are present in TAUOLA:
\begin{itemize}
\item CPC version~\cite{Jadach:1993hs};
It includes the dominant production mechanism, given by  $a_1 \to K* K$ and $a_1 \to \rho \pi$ for the axial-vector form factors and $\rho'\to (\rho\pi;K* K)$ for the vector form factors. The form factors are a product of the Breit-Wigner amplitudes for each separate resonance;  
\item RChL parametrization.
 In the case of the RChL approaches the vector current arises from  the Wess-Zumino term and the odd-intrinsic-parity amplitude~\cite{Dumm:2009kj}. The form factors receive contributions from the direct vertex, single-resonance and double-resonance mechanism production.
\end{itemize}

\begin{figure}\label{fig:prelKpiK}
\vspace{-0.3cm} 
\hspace{-0.5cm} 
\includegraphics[width = 0.25\textwidth]{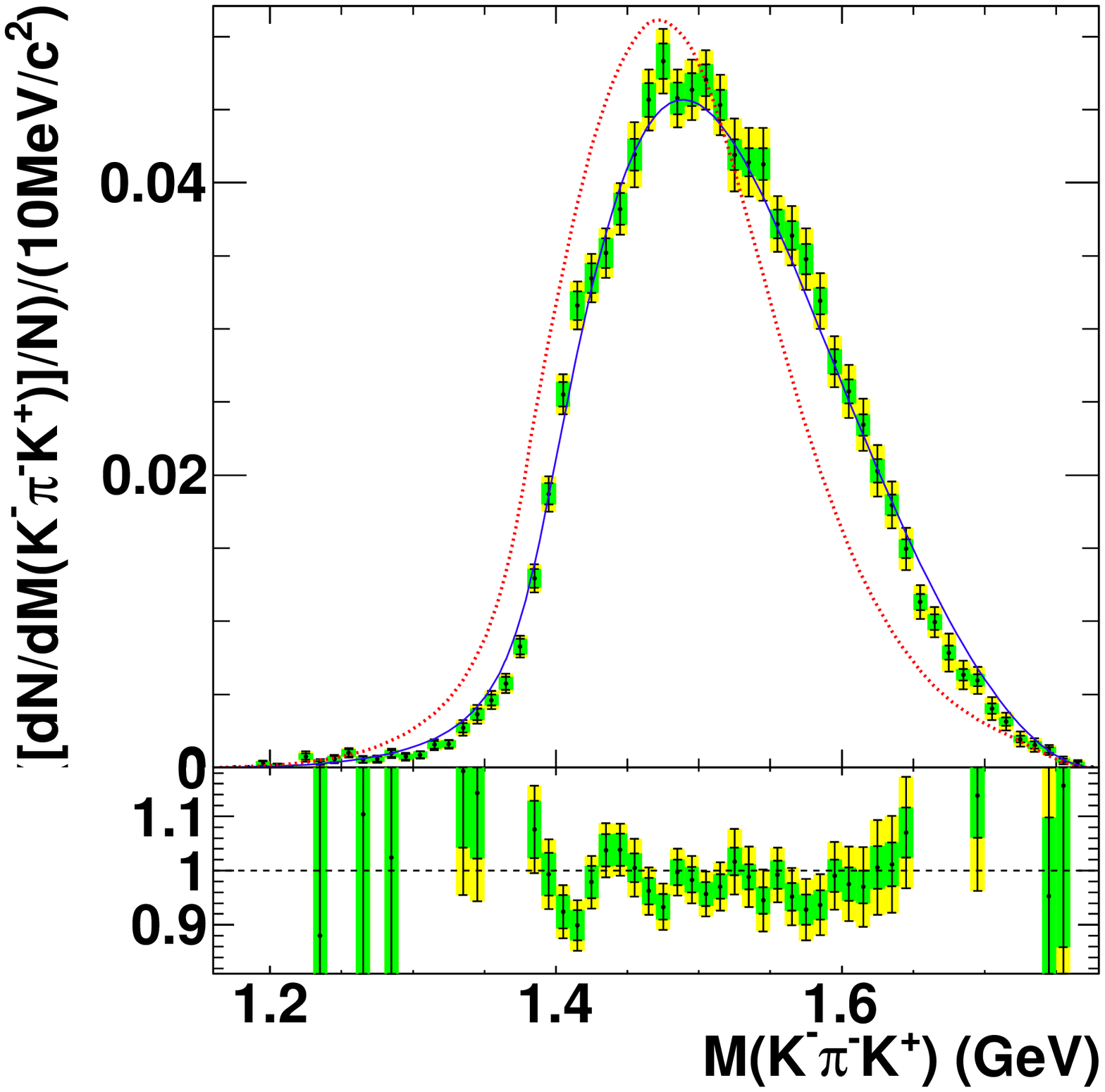}
\includegraphics[width = 0.25\textwidth]{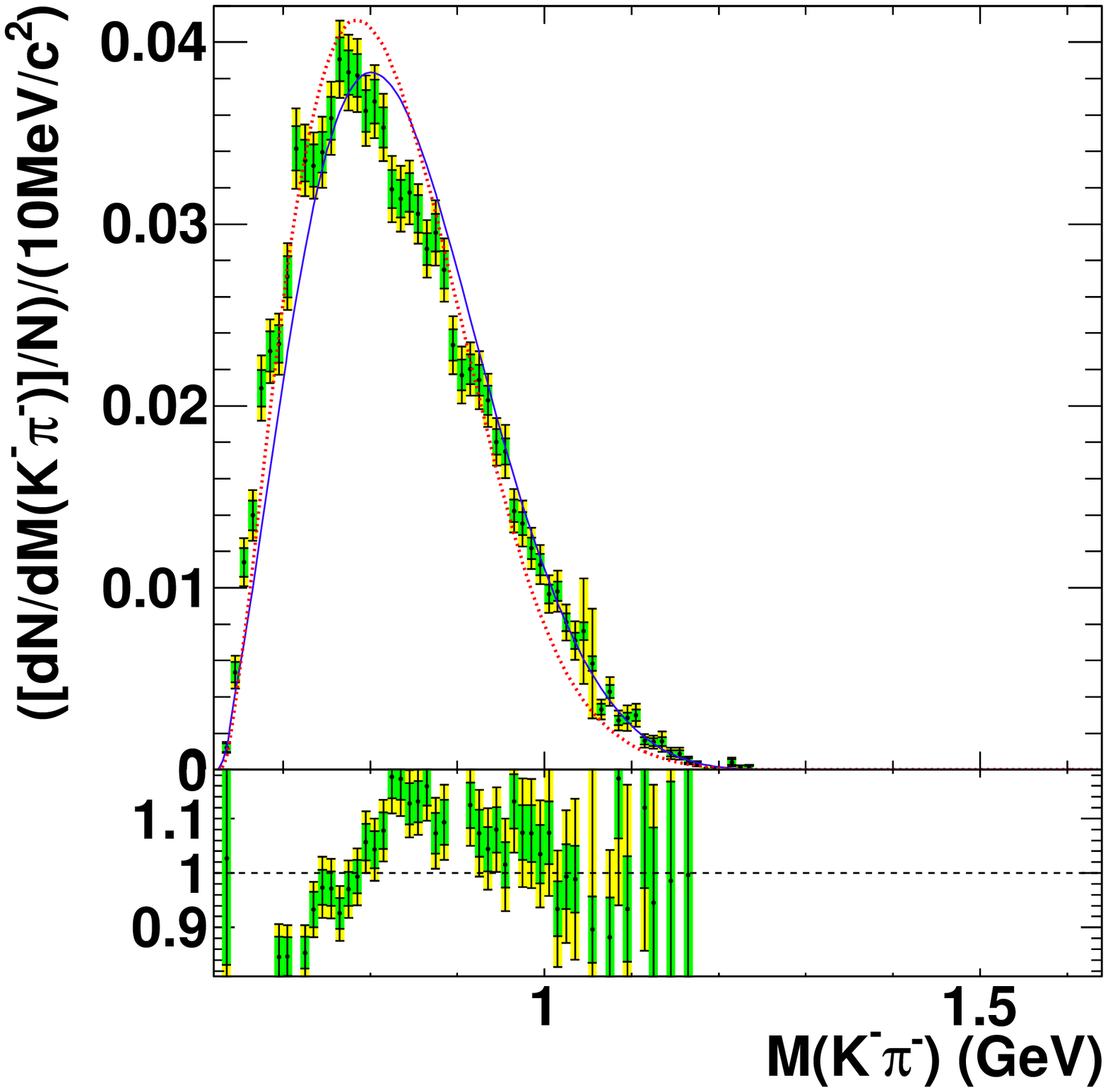}
\includegraphics[width = 0.25\textwidth]{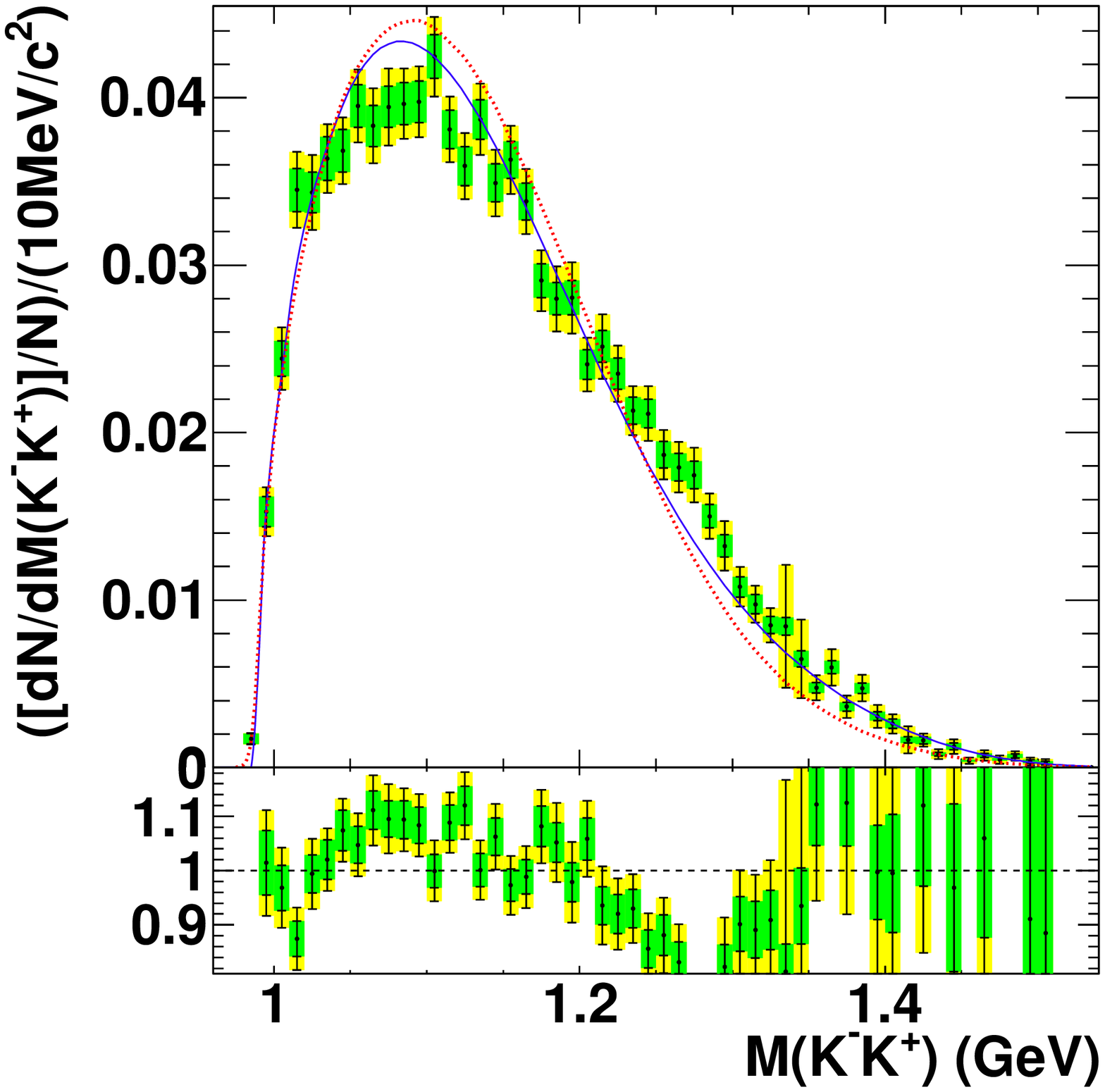}
\includegraphics[width = 0.25\textwidth]{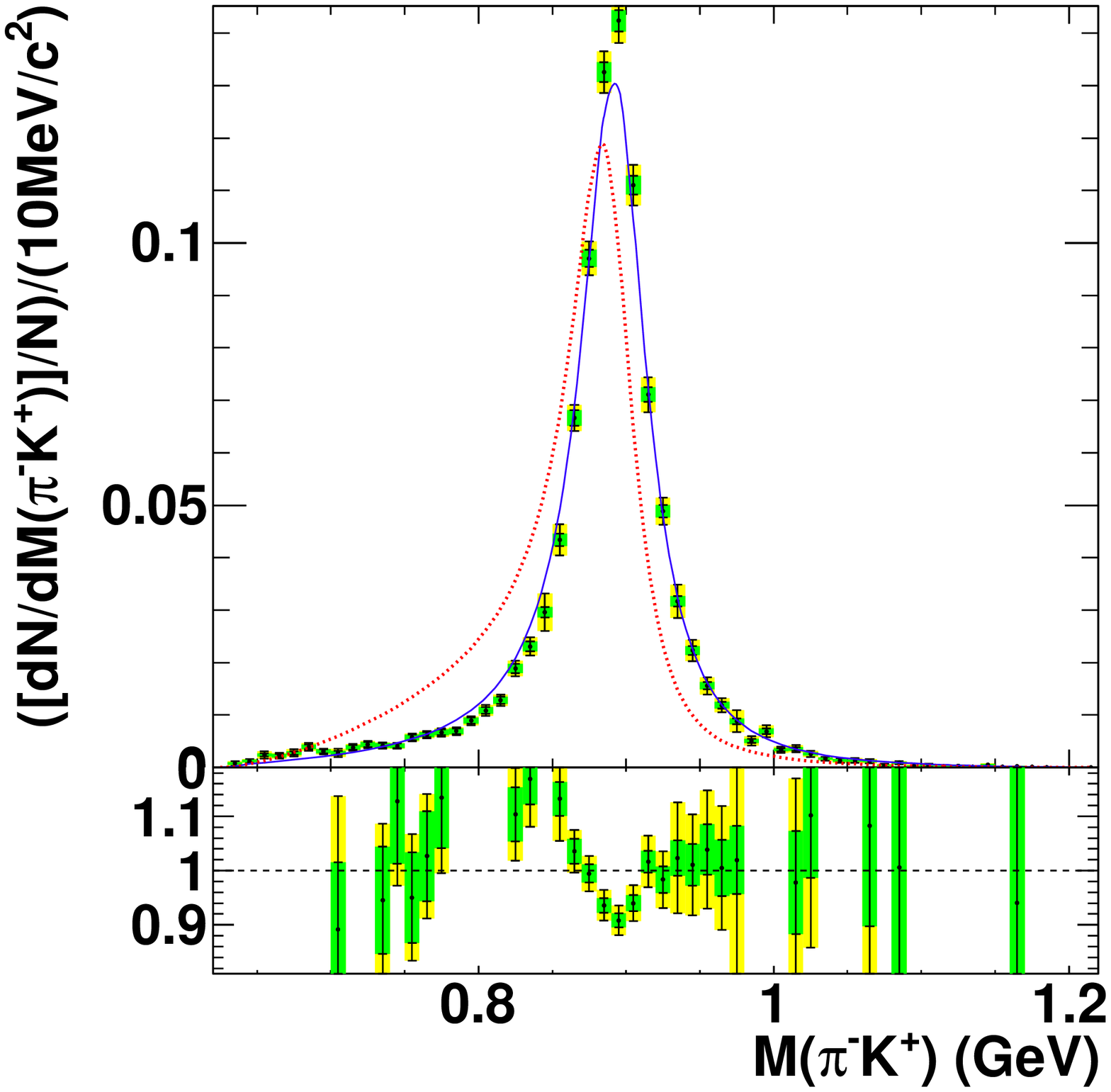}
\vspace{-0.3cm} 
\caption{The $\tau^- \to K^+ K^-\pi^-\nu_\tau$ decay 
invariant mass distribution of three and two meson system. For the description of the plots see Fig. 1.}
\end{figure}

The first fit  to the BaBar preliminary data~\cite{Nugent:2013ij} for  the two- and three-particle invariant mass spectra calculated on the base of the RChL parametrization  is presented in Fig.~\ref{fig:prelKpiK}.
It was carried out applying the generalized version of the fitting strategy used for the $\pi^-\pi^-\pi^+$ mode. Also the $a_1$ width was calculated only at the beginning of the fitting strategy and was not changed during the fit. 
An improved procedure might require a common fit of both $\pi^-\pi^-\pi^+$ and $K^+K^-\pi^-$ modes and work on this is in progress.  

\section{Conclusion}\label{sect:concl}

In this note we have reviewed theoretical parametrizations for the two- and three-meson $\tau$-decay modes included in the MC event generator TAUOLA.
In addition results of the fit for the invariant mass spectra of the two-pion decay mode to the Belle data and for the three-charged pion channels to the preliminary BaBar data have been  discused.

In the case of the two-pion channel the Belle data have been fitted with three parametrizations for the two-pion form factors.
The data have been reproduced with the Gounaris-Sakurai pion form factor parametrization while the RChL parametrization has failed.
Comparing the parametrization context we have concluded that the complex value of the resonance strength, used in the Gounaris-Sakurai parametrization, might tmimic the missing multiparticle loops. To check this idea we intend to evaluate the four-pion loops in the $\rho'$-resonance propagator that will be the object of future study.  

For the three-pion charged mode ($\pi^-\pi^+\pi^-$) we have fitted the BaBar preliminary data using the modified RChL parametrization. The corresponding theoretical approach is based on the Resonance Chiral Lagrangian with an additional  modification to the current to include the sigma meson.  As a result, we have improved agreement with the data by a factor of 
about eight compared with the previous results \cite{Shekhovtsova:2013rb}. Nonetheless, the model shows discrepancies in the high energy tail of the three pion invariant mass spectrum, that may be related with missing resonances, e.g. $a_1(1640)$, in the corresponding theoretical approach. 
We will come again on this point in future multidimensional analysis. 
The results on the numerical comparison between the TAUOLA three-pion parametrizations can be found in~\cite{Was:2015laa}.

Also we have presented the first results of the generalization of the fitting strategy to the case of an arbitrary three meson tau decay, specializing to the  $K^+K^-\pi^-$ decay mode. We have restricted ourselves to the pre-tabulated $a_1$ width approximation  and have not recalculated its value in the fit. This restriction will have to be removed in a common fit of both  $\pi^-\pi^-\pi^+$ and $K^+K^-\pi^-$ modes.   

TAUOLA upgrade is of the utmost importance in view of the forthcoming Belle-II project~\cite{Abe:2010gxa}.  The first physics run of the Belle-II project is planned in the fall of 2018. Both allowed and
forbidden tau decay modes in the Standard Model will be measured. Until 2022 it should record  a data sample fifty times larger than the BELLE experiment  and will require more precise theoretical modeling and further process simulation along the line of  this note. Therefore  the TAUOLA update will require both a more refined theoretical
approach for the hadronic mechanim production and the implementation of new hadronic modes in the code, for example, the $\eta$-meson modes~\cite{Escribano:2014joa}.

\section{Acknowledgements}
This research was supported in part by funds of the Foundation of Polish Science grant POMOST/2013-7/12, that is co-financed by the European Union, Regional 
Development 
Fund and by Polish National Science Centre under decisions  DEC-2011/03/B/ST2/00107.

\end{document}